\documentclass[prl,aps,showpacs,groupedaddress,twocolumn,floatfix]{revtex4}
\usepackage{amsmath}
\usepackage{bm}
\usepackage{graphics} 
\usepackage{dcolumn}
\begin{document}

\title{Relaxation  of  hole  spins  in  quantum  dots  via
  two-phonon processes}
\author{Mircea Trif$^{1}$, Pascal Simon$^{1,2}$, and Daniel Loss$^{1}$
}
\affiliation{$^{1}$Department   of  Physics,   University   of  Basel,
  Klingelbergstrasse      82,     CH-4056      Basel,     Switzerland}
\affiliation{$^{2}$ Laboratoire de Physique des Solides, CNRS UMR-8502 University Paris Sud, 91405 Orsay Cedex, France}

\date{\today}
\begin{abstract}

We investigate theoretically  spin relaxation  in heavy hole quantum dots in low
external  magnetic fields.  We  demonstrate that  two-phonon processes and spin-orbit interaction
are  experimentally  relevant   and  provide an explanation for the
recently  observed saturation  of the spin  relaxation rate  in  heavy hole
quantum  dots  with vanishing  magnetic fields.  We  propose  further
experiments  to  identify  the  relevant spin relaxation mechanisms  in  low  magnetic
fields.

\end{abstract}
\maketitle

In  the last  decade, remarkable  progress  has been made in  the
manipulation  and  control  of  the  spin  of  electrons  confined  in
semiconducting       nanostructures such as quantum
dots (QD) \cite{hanson:07}.
These achievements  pave the way  toward quantum spin  electronics and
may  lead  to  
spin-based  quantum computing \cite{loss:98}.  
In  the past
years, a new  candidate for a qubit state  has been attracting growing
interest: the spin of a heavy hole (HH) confined in a flat QD. In  a  bulk  
semiconductor  the HH  ($J_{z}=\pm3/2$)  and
light hole (LH) ($J_{z}=\pm1/2$) bands  are degenerate giving rise to
strong mixing and thus to strong  HH-spin relaxation. However, in a 2D
system the  HH and LH  bands are split  due to the  strong confinement
along the  growth direction \cite{winkler_book} implying  a significant
reduction of the HH spin relaxation via HH-LH mixing.

The relaxation ($T_1$) and decoherence ($T_2$) times of  a HH-spin  localized in  a QD
 are, like for electrons, determined by  the environment the
hole spin interacts  with:  the nuclear spin  bath  in the  QD  and the  lattice
vibrations (phonons).    
The former interaction is weaker for HHs than for electrons (due to the p-symmetry of  the hole) \cite{fischer:08,eble:08}.  More importantly, it is
of Ising type, making it ineffective
for HH-spins initialized along the growth direction  \cite{fischer:08}, as typically done in 
experiments \cite{karrai:08}. This is in contrast to electrons,
where the hyperfine interaction  is 
isotropic and  dominates the spin dynamics at low 
B-fields \cite{khaetskii:03,koppens2005,petta2005, imamoglu:07}.

The other relevant source of relaxation are phonons which 
couple to the hole spin through the spin-orbit  interaction (SOI) \cite{bulaev:05}. 
The predicted values \cite{bulaev:05}  for  the one-phonon  induced  
relaxation time  $T_1$ agree quite well with  data obtained
in  high B-fields \cite{finley:07}.  However,  for low B-fields
($B\sim1.5-3   {\rm   T}$)    and   high   temperatures   ($T>2   {\rm
  K}$),
a clear
deviation from  the one-phonon theory  has been  observed  \cite{finley:07}.
Furthermore,  recent experiments  on
optical pumping of HH-spins in QDs showed saturation of  $T_1$ for  very  low  or  
even vanishing  B-field \cite{karrai:08}.   
The   relaxation time   was   found   to   be  unusually long,
$T_1\approx0.1-1  {\rm  ms}$,  like
previously observed in high B-fields \cite{finley:07}.    
Both observations suggest other sources of relaxation, and the question arises what are they and
what are their observable consequences? 
The  answer to
this question is not only interesting by itself but also  relevant for using  HHs as 
 qubits.
In the following, we  show that  two-phonon processes are good candidates and even provide
a quantitative  explanation of the mentioned measurements at low B-fields \cite{finley:07,karrai:08}.  
Indeed, as we will see, these processes show weak or no B-field dependence, whereas
one-phonon relaxation rates 
vanish quickly with  $B\to 0$.
The importance of such two-phonon processes was noticed a long  time ago for
 electron spins  in silicon-donors \cite{abrahams:57} and
rare-earth ions \cite{capellmann:89}, while for electrons in QDs,  they have been recently analyzed and shown to be negligible
compared to nuclear spin effects 
\cite{pablo:06}. 

 To describe a  HH confined to a QD  and interacting with the
 surrounding phonon bath, we start with the following Hamiltonian
\begin{equation}
H_{h}=H_0+H_Z+H_{SO}+H_{h-ph}+H_{ph},\label{hamiltonian}
\end{equation}   
where $H_0=p^2/2m^*+V(\bm{r}),\label{orbital}$ is the dot Hamiltonian,
$V(\bm{r})\equiv  m^*\omega_0^2r^2/2$  is  the  confinement  potential
which   is   assumed   to    be   harmonic, with $m^*$ being the HH mass.   The   second   term   in
Eq.  (\ref{hamiltonian})  is  the  Zeeman  energy  of  the HH
(pseudo-) spin  $H_Z=g\mu_B\bm{B}\cdot\bm{\sigma}/2\label{Zeeman}$,  with
$\bm{B}$ being the magnetic field and $\bm{\sigma}$ the  Pauli
matrices  for the HH spin defined in  the  $J_z=\pm3/2$  subspace.   The  third  term
represents the spin-orbit Hamiltonian, which,  for well separated  HH-LH bands (flat dots), reads \cite{bulaev:05}
\begin{eqnarray}
H_{SO}&=&\beta p_-p_+p_-\sigma_++h.c. \label{SOI}
\end{eqnarray}
This  Hamiltonian represents  the  effective Dresselhaus SOI   (restricted to the HH subspace) due to  bulk
inversion   asymmetry   of   the   crystal \cite{bulaev:05},   where
$p_{\pm}=p_x\pm  ip_y$, $\bm{p}=-i\hbar\bm{\nabla}-e\bm{A}(\bm{r})$, $\bm{A}(\bm{r})=(-y,x,0)B/2$, and $\sigma_{\pm}=\sigma_x\pm  i\sigma_y$. We
note  that in  Eq. (\ref{SOI})  we have  neglected the  Rashba SOI and other possibly linear-in-k but small SOI terms
 \cite{bulaev:05}.    The  fourth   term  in   Eq.  (\ref{hamiltonian})
represents the interaction  of the HH charge with  the phonon field,
{\it i.e.} $H_{h-ph}=\sum_{\bm{q}j}M_{\bm{q}j}X_{\bm{q}j}$ with
\begin{eqnarray}
M_{\bm{q}j}=\frac{F(q_z)e^{i\bm{q}\cdot\bm{r}}}{\sqrt{2\rho_c\omega_{\bm{q}j}}}\left[e\beta_{\bm{q}j}-i(\Xi_0\,\bm{q}\cdot\bm{d}_{\bm{q}j}-\Xi_z\,q_zd_{\bm{q}j}^z)\right],\label{HHphonon}
\end{eqnarray}
and
$X_{\bm{q}j}=\sqrt{\hbar/\omega_{\bm{q}j}}(a^{\dagger}_{-\bm{q}j}+a_{\bm{q}j})$,
where $\bm{q}$ is  the phonon wave-vector, with $j$  denoting the acoustic
branch,  $\omega_{\bm{q}j}=c_jq$  the phonon energy, with $c_j$  the speed of sound in
the $j$-th  branch, $\bm{d}_{\bm{q}j}$  the polarization unit vector, $\rho_c$  the sample density (per unit volume),   and $e\beta_{\bm{q}j}$  
the  piezoelectric electron-phonon
coupling and $\Xi_{0,z}$ the  deformation potential constants \cite{bulaev:05}. 
   The form  factor $F(q_z)$  in  Eq. (\ref{HHphonon})
equals unity for $|q_z|\ll d^{-1}$  and  zero for $|q_z|\gg d^{-1}$,
with  $d$  being  the  dot  size  in  the (transverse) z-direction. 
The   last  term  in  Eq.   (\ref{hamiltonian}) describes the free phonon bath.

In the  following, we  analyze the  effect of the  phonons  on the
 HH spin, both in the low and high temperature regimes.
The phonons do not couple directly  to the spin, but
the SOI  plays the  role of the  mediator of an  effective spin-phonon
interaction. Under the realistic assumption  that the level splitting in the dot is much larger than the HH-phonon interaction, we can treat  $H_{h-ph}$ in perturbation theory.

Let  us  define the  dot  Hamiltonian  $H_d\equiv H_0+H_Z+H_{SO}$  and
denote  by  $|n\sigma\rangle$, the  eigenstates  of  $H_d$ (where  $n$
labels the  orbital and $\sigma$ the  spin states). 
The product states $|n\rangle|\sigma\rangle$  are then the eigenstates of
$H_d$ in the absence of SOI ({\it i.e.} for $H_{SO}=0$).
These states are formally connected to $|n\sigma\rangle$
by an exact Schrieffer-Wolff (SW) transformation \cite{pikus_book,vitaly:04}, i.e., 
$|n\sigma\rangle=e^{S}|n\rangle|\sigma\rangle$, 
 where $S=-S^{\dagger}$ is the SW generator  
 and can  be found in
 perturbation  theory   in  SOI.    After  this
 transformation,  any operator  $A$  in the  old  basis transforms  as
 $A\rightarrow\widetilde{A}=e^{S}Ae^{-S}$ in the new basis (e.g.,
 $H_{d}\rightarrow\widetilde{H}_d$, $H_{h-ph}\rightarrow
 \widetilde{H}_{h-ph}$, {\it etc.}).

Let us now derive the effective spin-phonon interaction 
under  the  above  assumptions.  To  do  so,  we  perform  another  SW
transformation of the total HH Hamiltonian $H_{h}$ to obtain an effective
Hamiltonian  $H_{\it eff}=e^{T}H_h  e^{-T}$, where  $T=-T^{\dagger}$  is  chosen  such  that  it  diagonalizes  the
HH-phonon Hamiltonian $\widetilde{H}_{h-ph}$ in the eigenbasis of 
$H_d$. In lowest  order in  ${H}_{h-ph}$, we
obtain
$T\approx   \widetilde{L}_d^{-1}\widetilde{H}_{h-ph}$,
where the Liouvillean is defined as $\widetilde{L}_{d}A=[\widetilde{H}_{d},A]$,     $\forall    A$,
and diagonal terms of $ {H}_{h-ph}$ are to be excluded.
In 2nd  order in ${H}_{h-ph}$, we obtain then the effective spin-phonon Hamiltonian
\begin{eqnarray}
H_{s-ph}&=&\bm{\sigma} \cdot \sum_{\bm{q}j,\bm{q}'j'} \big[ \delta_{\bm{q}j,\bm{q'}j'} \bm{C}^{(1)}_{\bm{q}j}\,X_{\bm{q}j}+\bm{C}^{(2)}_{\bm{q}j,\bm{q}'j'} X_{\bm{q}j}X_{\bm{q}'j'}\nonumber\\&+&\bm{C}^{(3)}_{\bm{q}j,\bm{q}'j'}\,\big(P_{\bm{q}j}\,X_{\bm{q}'j'}-P_{\bm{q}'j'}\,X_{\bm{q}j}\big)\big],\label{spinphonon}
\end{eqnarray}
with
$\bm{\sigma}\cdot\bm{C}^{(1)}_{\bm{q}j}\!\!\!=\!\!\!\langle0|\widetilde{M}_{\bm{q}j}|0\rangle$,
$\bm{\sigma}\cdot\bm{C}^{(2)}_{\bm{q}j,\bm{q}'j'}\!\!=\!\!\!\langle0|[\widetilde{L}_d^{-1}\widetilde{M}_{\bm{q}j},\widetilde{M}_{\bm{q}'j'}]|0\rangle$,
$\bm{\sigma}\cdot\bm{C}^{(3)}_{\bm{q}j,\bm{q}'j'}\!\!\!\!=\!\!\!\!\langle0|[\widetilde{L}_d^{-1}\widetilde{M}_{\bm{q}j},\widetilde{L}_d^{-1}\widetilde{M}_{\bm{q}'j'}]|0\rangle$,
$P_{\bm{q}j}=i\sqrt{\hbar\omega_{\bm{q}j}}(a^{\dagger}_{-\bm{q}j}-a_{\bm{q}j})$
is the  phonon field momentum operator, and $|0\rangle$ is the orbital ground state.   In Eq.   (\ref{spinphonon}) we have
neglected  spin-orbit corrections to the  energy
levels, being 2nd order in SOI.  Note
that for vanishing  magnetic field $\mathbf B\to 0$ the only
term  which survives  in  $H_{s-ph}$  is the  last one
since this  is the only one which preserves  time-reversal invariance and thus gives rise to zero field relaxation (ZFR) \cite{abrahams:57,capellmann:89,pablo:06}.
Quite remarkably,   this term is  1st order in SOI, whereas for  electrons it is only 2nd order \cite{pablo:06}.  
This is  one of the main reasons why eventually two-phonon processes are much more effective for HHs than for electrons.

We  now assume  the  orbital confinement  energy $\hbar\omega_0$  much
larger than  the SOI, {\it i.e.}  $||H_0||\gg ||H_{SO}||$, so  that we can
treat  the   SOI  in  perturbation   theory.  We  consider   also  the
$\bm{B}$-field to be  applied perpendicularly to the dot  plane (as in Refs.~\cite{finley:07,karrai:08}).  We limit our  description to first order effects in
SOI. The  SW-generator $S$  can be  written as
$S=S_+\sigma_- -h.c.$, and we then find
\begin{eqnarray}
\!\!\!\!&S_{+}&\!\!\!\!=A_1p_+p_-p_++A_2[p_+p_-P_+-(p_+P_--P_+p_-)p_+]\,\,\,\nonumber\\
&\!\!\!+&\!\!\!\!\!A_4P_+P_-P_++A_3[(p_+P_--P_+p_-)P_++P_+P_-p_+]. \,\,\,\,
\end{eqnarray}
Here, $A_i\equiv      A_i(\omega_Z,\omega_c)$ with $\omega_Z=g\mu_BB/\hbar$ and $\omega_c=eB/2c$.  For
$\omega_Z,\omega_c\ll\omega_0$, we obtain $A_1\approx-(7\beta/9\hbar)
((\omega_Z+\omega_c)/\omega_0^2)$,
$A_{2}\approx-(\beta/3\hbar)(\omega_c/\omega_0^2)$,
$A_3\approx-(2\beta/9\hbar)(\omega_c^2(\omega_c+\omega_Z)/\omega_0^4)$,
and       $A_4\approx(2\beta/3\hbar)(\omega_c^3/\omega_0^4)$,      while
$P_{\pm}=P_x\pm                       iP_y$                       with
$P_{x(y)}=-i\hbar\nabla_{x(y)}\pm(m^*\omega_0^2/\omega_c) y(x)$.
After somewhat tedious calculations,  we   obtain  analytic
expressions   for    $\mathbf   C^{(i)}=(C^{(i,x)},C^{(i,y)},0)$  
occurring in Eq. (\ref{spinphonon}).
We     give     below     only     the    exact     expression     for $i=3$,
 the  rest being  too lengthy to
be displayed here:
\begin{eqnarray}
&C^{(3,x/y)}_{\bm{q}j,\bm{q}'j'}&=\pm M_{\bm{q}j}^{\bm{q}'j'}\frac{m^*\beta e^{-b^2/4}   }{3\lambda_d\hbar\omega_0^2}\mathcal{F}(\bm{b}\cdot\bm{b}')
\nonumber\\ \!&\times&
     \!\!\!\!\Big(b_y^2       b_x^{'}-b_y^{'2}       b_x
\pm (b_x-b_x')(2b_yb_y'+3b_xb_x')\Big),
\end{eqnarray}
where
\begin{eqnarray}
\mathcal{F}(\bm{b}\cdot\bm{b}')&=&\frac{1}{(\bm{b}\cdot\bm{b}')^2}\big(e^{-\bm{b}\cdot\bm{b}'/2}-\bm{b}\cdot\bm{b}'/2\big)\nonumber\\&\times&\big(\gamma+\log(\bm{b}\cdot\bm{b}'/2)+\Gamma(0,\bm{b}\cdot\bm{b}'/2)\big).
\end{eqnarray}
Here,
$M_{\bm{q}j}^{\bm{q}'j'}=(F(q_z)F(q_z')\hbar/2\rho_c\sqrt{\omega_{\bm{q}j}\omega_{\bm{q}'j'}})(\Xi_0\,\bm{q}\cdot\bm{d}_{\bm{q},j}-\Xi_z\,q_zd_{\bm{q}j}^z)(\Xi_0\,\bm{q}'\cdot\bm{d}_{\bm{q}'j'}-\Xi_z\,q_z'd_{\bm{q}'j'}^z)$
and     $\bm{b}=\bm{q}\lambda_d$,     where
$\lambda_d$  is  the dot-diameter.  We  have  also  introduced
$\gamma\approx 2.17$  the Euler  constant and $\Gamma(s,x)$   the incomplete  gamma function. We  note that  $C^{(1,2)}\propto B$,  so that these  two terms vanish  with vanishing B-field. 

Let  us now  analyze the  relaxation of  the spin  induced by  all the
phonon   processes    in   the   spin-phonon    Hamiltonian   in   Eq.
(\ref{spinphonon}).    We   first    mention   that   all   terms   in
Eq. (\ref{spinphonon}) can  be cast in a general  spin-boson type of
Hamiltonian
$H_{s-b}^{p}=(1/2)g\mu_B\delta{\bm{B}}^{p}(t)\cdot\bm{\sigma}, p=1,2,3$,
with  the  corresponding identification  of  the fluctuating  magnetic
field  terms  $\delta{B}_j(t)$   from  Eq.   (\ref{spinphonon})  (e.g.
$\delta{\bm{B}}^1(t)\sim         \bm{C}^{(1)}_{\bm{q}j}X_{\bm{q}j}$).

Within  the Bloch-Redfield  approach,  the relaxation  rate $\Gamma\equiv 1/T_1$ can be expressed as
$\Gamma=\sum_{i=x,y}\left[J_{ii}(E_Z/\hbar)+J_{ii}(-E_Z/\hbar)\right]$.
 The    correlation     functions    $J_{ij}$    are     defined    by
 $J_{ij}(\omega)=(g\mu_B/2\hbar)^2\int_{0}^{\infty}\!\!dt   e^{-i\omega
   t}\!\!<\delta{B}_{i}(0)\delta{B}_{j}(t)>$,  where  $<\dots>$ denotes
 the  average  over  the  phonon   bath,  assumed  to  be  in  thermal
 equilibrium at temperature T.  The relaxation time associated with the three types of
 spin-phonon processes in Eq. (\ref{spinphonon}) is $\Gamma=\sum_{i=1,2,3}\Gamma^{(i)}$ with
\begin{eqnarray}
\Gamma^{(1)}&=&\frac{4\pi}{\hbar}\sum_{\bm{q}j}|\bm{C}^{(1)}_{\bm{q}j}|^2\left(n(\omega_{\bm{q}j})
+\frac{1}{2}\right)\delta(E_Z-\hbar\omega_{\bm{q}j}),
\label{single_relax}\nonumber\\
\Gamma^{(m)}&\simeq&\frac{8\pi}{\hbar}\sum_{\bm{q}j,\bm{q}'j'}|\bm{C}^{(m)}_{\bm{q}j,\bm{q}'j'}|^2(\omega_{\bm{q}j}\omega_{\bm{q}'j'})^{m-2}n(\omega_{\bm{q}j})\nonumber\\&\times&\left(n(\omega_{\bm{q}'j'})+1\right)\delta(\hbar\omega_{\bm{q}j}-\hbar\omega_{\bm{q}'j'})\label{two_relax},
\end{eqnarray}
where  $n(\omega)=1/(\exp{(\omega/k_BT)}-1)$ is  the  Bose factor and $m=2,3$ correspond to B-dependent and B-independent two-phonon rates, {\it resp}.
We remark that in Eq.  (\ref{two_relax}) we have neglected some irrelevant
processes  in the limit  of low-B  field \cite{khaetskii:01}. Also, for $B$-fields  perpendicular to the dot plane the decoherence time satisfies $T_2=2T_{1}$  for  one-and two-phonon processes  since the spin-phonon fluctuations $\delta{{\bm B}}_j\perp{\bm B}$ \cite{vitaly:04,bulaev:05}. 

Note that for two-phonon processes the
single phonon-energies do not need to match the
Zeeman energy separately (as opposed to one-phonon processes),
so that there is only a weak dependence on the B-field left which
comes from the effective spin-phonon coupling itself.

\begin{figure}[t]
\scalebox{0.34}{\includegraphics{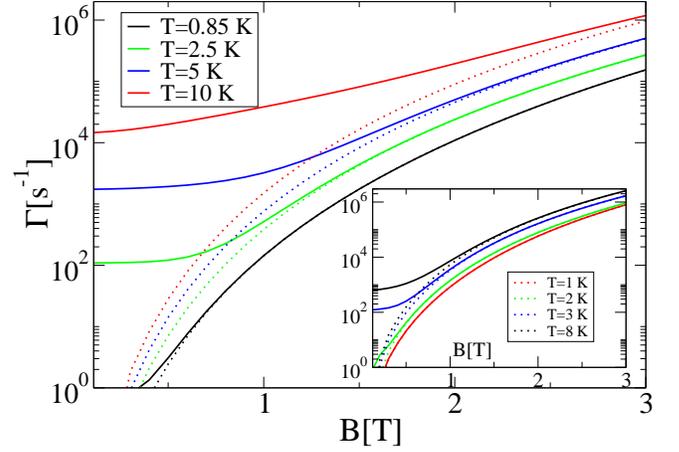}}
\caption{The heavy-hole spin relaxation  rate $\Gamma$ for InAs QDs (GaAs QDs in the inset) as a function  of magnetic field
  $B$ for different temperatures T. The full curves represent the 
  rate  due to one-  and two-phonon processes,  {\em i.e.}
  $\Gamma=\sum_{i=1}^3\Gamma^{(i)}$   as  defined  in
  Eq. (\ref{two_relax}) for different temperatures T, while the dotted lines present the one-phonon rate $\Gamma^{(1)}$.}
\label{singlerelax}
\end{figure}

\begin{figure}[t]
\scalebox{0.34}{\includegraphics{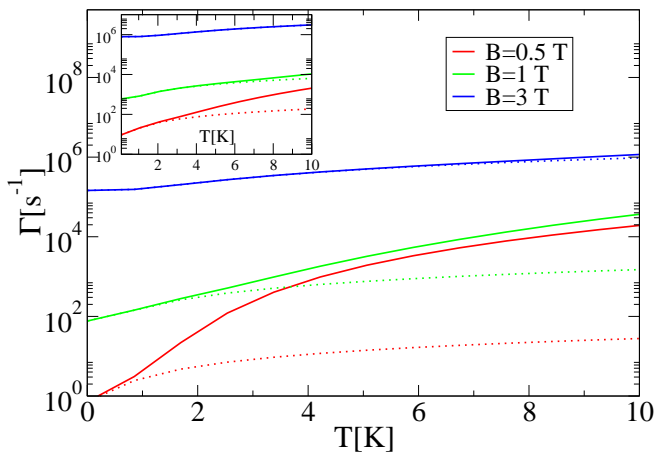}}
\caption{The heavy-hole spin relaxation  rate $\Gamma$ in Eq. (\ref{two_relax}) for InAs QDs (GaAs QDs in the inset) as a  function of temperature T  for
  different B-field values. For finite $B$-field,  $\Gamma$ saturates at low temperatures due to one-phonon processes.}
\label{relax1d}
\end{figure}

 In  Figs.
\ref{singlerelax} and \ref{relax1d}, we plot the  phonon spin-relaxation rate $\Gamma$ as
a function of the B-field and of temperature, resp., for InAs and GaAs quantum dots. Fig. \ref{singlerelax} shows a clear saturation of
$\Gamma$ at low magnetic fields which is due to two-phonon processes, while Fig. \ref{relax1d} shows the known saturation at low temperatures due to one-phonon processes \cite{bulaev:05}.

  For these plots, we  used the following HH  InAs QDs (labeled by A) \cite{Kroner1:08,Kroner2:08} and GaAs QDs (labeled by B)
parameters \cite{bulaev:05}: $\Xi_0=1.9\,  {\rm eV}$,  $\Xi_z=2.7 {\rm eV}$,
$c_t^{A}=2.64\cdot10^{3}\,{\rm m/s}$ ($c_t^{B}=3.35\cdot10^{3}\,{\rm
  m/s}$), $c_l^{A}=3.83\cdot10^3\,{\rm m/s}$
($c_l^B=4.73\cdot10^3\,{\rm m/s}$),
$\rho_c^{A}=5.68\cdot10^{3}\,\rm{kg/m^3}$
($\rho_c^{B}=5.3\cdot10^{3}\,\rm{kg/m^3}$), $m^*_{A}=0.25  m_e$
($m^*_{B}=0.14  m_e$), $g^{A}=1.4$ ($g^{B}=2.5$), and we assume
$\lambda_d=3 {\rm nm}$ ($\hbar\omega^{A}_0=35\, {\rm meV}$,
$\hbar\omega^{B}_0=60\,{\rm meV}$)  and   $d=3\,\rm{nm}$   (dot
height). Also, $\beta_{A}\approx 2.1\cdot10^5\,{\rm m/s}$ and $\beta_{B}\approx 4.6\cdot10^4\,{\rm m/s}$. From
Fig. \ref{singlerelax}   we can  infer  that the two-phonon  processes
become dominant for magnetic  fields $B<2 \rm{T}$ ($B<0.5 \rm{T}$) and for
temperatures $ T>2 \rm{K}$ ($T>3 {\rm K }$) for InAs (GaAs) QDs. These  estimates for the relaxation
rates due to one- and two-phonon processes  are comparable to the ones recently  measured
in Refs. \cite{finley:07,karrai:08}, thus providing a
reasonable explanation for these measurements.  Note that, in contrast
to the HH case,  the relaxation time for electrons shows no deviation
from the one-phonon time (or saturation) with decreasing
$B$-field \cite{amasha:06}. 

Next, we provide explicit expressions of the relaxation rates for low and high temperature limits.
The rates $\Gamma^{(i)}$ can be written  as 
\begin{equation}
\Gamma^{(i)}=\delta_i\sum_{m=0}^{r_i}\frac{\omega_Z^{r_i-m}\omega_c^m}{\omega^{r_i}_0}F_{i}^{(m)}\left(t\right),\label{rate_explicit}
\end{equation}
where $\delta_1\approx 2\pi(\hbar^4 eh_{14}^2\beta^2/\kappa^2m_{h}\lambda_d^6\rho_cc_l^5)$, $\delta_2\approx \pi (m_{h}^4\beta^2\Xi_0^4/\hbar^2\lambda_d^5\rho_c^2 c_l^3)$, $\delta_3\approx\pi (m_h^6\beta^2\Xi_0^4/\hbar^4c_l\lambda_d^3\rho_c^2)$,  $r_1=5$, $r_2=2$, $r_3=0$, and $t=k_BT/E_{ph}$ with  $E_{ph}\equiv\hbar c_l/\lambda_d$.
\begin{table}[h]
\centering
\begin{tabular}{c c c c c c c}
\hline
\hline
& $F_1^{(0)}$ & $F_1^{(1)}$  & $F_2^{(0)}(t)$ & $F_2^{(1)}(t)$ & $F_2^{(2)}(t)$ & $F_3(t)$  \\[0.3ex]
\hline
\\[0ex]
 $t\ll1$ & $0.004$ & $0.015$ & $10^8t^{13}$ & $10^7 t^{13}$ & $5\cdot10^6t^{13}$ & $10^9t^{15}$ \\[1ex]
\\[0ex]
$t\gg1$ & $0.08\frac{t}{\omega_Z}$ & $0.03\frac{t}{\omega_Z}$ & $10^2t^{2}$ & $10^2 t^{2}$ & $30t^{2}$ & $0.3\, t^{2}$ \\[1ex]
\hline
\hline
\end{tabular}
\caption{The asymptotic values for $F_i^{(m)}(t)$.}
\label{tab1}
\end{table}
The functions  $F_{i}^m(t)$ depend on the ratios $t=k_BT/E_{ph}$,  $d/\lambda$, and $c_l/c_t$. In Table ~\ref{tab1}
we list the asymptotic (scaling) expressions for  $F_i^{(m)}(t)$ in low B-fields $\omega_{c,Z}\ll \omega_0$ for low  
($t\ll1$) and  high   ($t\gg 1$) temperatures. 
We note that $F_{1}^{(1)}(t)\approx F_1^{(2)}(t)$ in both regimes, and $F_{1}^{(3,4,5)}\equiv0$. 

Using Eq.~(\ref{rate_explicit}) and Table ~\ref{tab1}  we can write for the two-phonon rates, say, for InAs QDs
\begin{eqnarray}
\Gamma^{(2)}\!=\!\!\delta_2\left\{\begin{array}{ll}
\!\!10^7\left(10\,\frac{\omega_Z^2}{\omega^2_0}+\frac{\omega_Z\omega_c}{\omega^2_0}+0.5\,\frac{\omega_c^2}{\omega^2_0}\right)\,T^{13}, & T\ll E_{ph}\\
\!\!10^2\left(\,\frac{\omega_Z^2}{\omega^2_0}+\,\frac{\omega_Z\omega_c}{\omega^2_0}+0.3\,\frac{\omega_c^2}{\omega^2_0}\right)\,T^2, & T\gg E_{ph}
\end{array}\right.\nonumber
\end{eqnarray}

\begin{eqnarray}
\Gamma^{(3)}\!=\,\delta_3\left\{\begin{array}{ll}
\!10^9\,T^{15}, & T\ll E_{ph}\\
\!0.3\,T^2, & T\gg E_{ph}.
\label{asymptotrates}
\end{array}\right.
\end{eqnarray}

 From Eqs.~(\ref{asymptotrates})  we find  that for $T<2 \rm{K}$ and for  $B>0.5 \rm{T}$ the one-phonon processes dominate the relaxation rate $\Gamma$. On the other hand, for low B-fields ($0.1 \rm{T}<B<1 \rm{T}$) and finite temperatures ($T>2\rm{K}$) the two-phonon processes will give the main contribution to  $\Gamma$, see Fig. \ref{relax1d}. The main phonon processes could be identified experimentally by analyzing the temperature dependence of $\Gamma$, scaling  as $\Gamma\sim T$ for one-phonon processes and as $\Gamma\sim T^2$ for two-phonon processes. Also, the saturation of $\Gamma$ in vanishing B-field is a clear indication of two-phonon processes.  Note that the strong enhancement of the two-phonon HH spin relaxation arises because (i) the rate is  2nd order in SOI (whereas for electrons it is 4th order) and (ii) the effective mass for HHs is much larger than that for electrons.

In order  to compute  $\Gamma^{(2,3)}$, we took into  account only the
contribution from  the deformation potential since  this dominates the two-phonon  relaxation for  $T/E_{ph}>0.1$ and $\omega_Z,\omega_c\ll\omega_0$. For the evaluation of $\Gamma^{(1)}$ instead, we considered both the piezoelectric and deformation potential  contributions, both of them being important for $B$ and T considered here.  
Surprisingly, we found that the ZFR rate $\Gamma^{(3)}$ increases when decreasing the
dot size  as $\Gamma^{(3)}\sim \lambda_d^{-1}$, while the other two rates decrease with decreasing the dot size as $\Gamma^{(1)}\sim\lambda_d^4$ and $\Gamma^{(2)}\sim\lambda_d$.  
This behavior  strongly  differs  from  the  electronic case  where  the  ZFR
mechanism is efficient for rather large dots \cite{pablo:06}.  

Interestingly, the present results do not change much if the $B$-field is tilted with respect to the QD plane. The $g$-factor for HHs  is strongly anisotropic with $g_{\parallel}\ll g_{\perp}$ so that one can neglect the in-plane Zeeman splitting. This implies performing the substitution $\omega_{c,Z}\rightarrow \omega_{c,Z}\cos{\theta}$ in above results, with $\theta$ being the angle between the $B$-field and the $z$-direction. This will lead to a reduction of the $B$-dependent rates ($\Gamma^{(1,2)}$), while the ZFR ($\Gamma^{(3)}$) being independent of $B$ remains the same.

In  conclusion,  we  have  shown  that
two-phonon processes give rise to a strong relaxation of  the HH spin  in a flat quantum dot.   This time  is predicted to  be in the  millisecond range,
comparable  to  the one  measured  in  recent  experiments on  optical
pumping of a HH spin in QDs \cite{karrai:08}.  Though other sources of
 relaxation  are  not excluded,  a  careful  scaling  analysis of  the
measured relaxation time with the magnetic field and/or the temperature should allow one to
 identify  the two-phonon process as  the leading relaxation
mechanism for the heavy-hole spin localized in small QDs.

We thank 
D. Bulaev, J. Fischer, V. Golovach, and  D. Stepanenko for useful discussions. This
work was supported  by the Swiss NSF, NCCR  Nanoscience, and JST
ICORP.


\end{document}